# Agent Based Negotiation using Cloud - an Approach in E-Commerce

Amruta More[1], Sheetal Vij[1], Debajyoti Mukhopadhyay[2],

[1] Department of Computer Engineering ,
[2] Department of Information Technology,
Maharashtra Institute of Technology, Pune- 411038, India,
{moreamruta930, sheetal.sh, debajyoti.mukhopadhyay} @gmail.com

**Abstract.** 'Cloud computing' allows subscription based access to computing. It also allows storage services over Internet. Automated Negotiation is becoming an emerging, and important area in the field of Multi-Agent Systems in E-Commerce. Multi-Agent based negotiation system is necessary to increase the efficiency of E-negotiation process. Cloud computing provides security and privacy to the user data and low maintenance costs. We propose a Negotiation system using cloud. In this system, all product information and multiple agent details are stored on cloud. Both parties select their agents through cloud for negotiation. Agent acts as a negotiator. Agents have user's details and their requirements for a particular product. Using user's requirement, agents negotiate on some issues such as price, volume, duration, quality and so on. After completing negotiation process, agents give feedback to the user about whether negotiation is successful or not. This negotiation system is dynamic in nature and increases the agents with the increase in participating user.

## 1  Introduction

In business negotiation two or more parties come together to find mutually agreeable contractual decision. For negotiation process both parties must show their interest, thus, negotiation can be complicated and lengthy process. When all parties have to come to final decision, negotiation will be stopped. In negotiation, each individual aim to achieve the best possible outcome for their organization.

Negotiator is an individual representing an organization which listens to all the parties' decision carefully and takes his own decision which gives profit to his organization. In negotiation process organization profit depends on organization's negotiator so that negotiator needs to understand situation and all other organization's negotiator. Negotiator must know how to negotiate well to successfully close deals, avoid conflicts, and establish better relations among the other organization's negotiators making the organization a better place to work. For successful negotiation individuals or negotiator must learn to compromise and stop finding faults in each other.

Today cloud computing is widely used and is becoming a popular technology. Cloud is a remote server, where the user can store their data and access the data remotely whenever it's required. Cloud computing provides security and privacy to the user data. User has no burden in maintaining huge amount of data stored on cloud. Cloud computing is sometimes referred to as "on-demand resources" and is usually based on pay-per-use basis. If business owner requires more space as compared to his previous space on cloud, owner can easily request for additional data storage on the cloud. Also owner can easily request for additional bandwidth, processing speed, and additional licenses. Using cloud computing, user can access information from any device like desktop, minicomputer, mobile etc. anywhere and anytime.

Amazon, Microsoft, Openstack, Google all these are the cloud providers. Google Apps is one example of cloud.

In Agent based negotiation system, we propose a system for negotiation between a provider and a consumer using cloud. In this system, all product information and multiple agent details are stored on the cloud. Both provider and consumer will select their agents through cloud for negotiation. An agent acts as a negotiator. Agent has user's details and their requirements for a particular product. Using user's requirement, agents negotiate on certain features or issues.

## 2  Features of Cloud for E-Negotiation

1. **Rapid Scalability:** Cloud computing has the ability of scaling the resources both ways for the consumers and as per the need. Cloud is infinite and one can buy the computing power as per the need. Negotiation system is dynamic, so that if more data storage is required, it can be easily made available by cloud.
2. **Security:** It is the core feature of cloud computing. Security is much stricter in cloud computing. Data is shared within a server hence, the provider must ensure that each account is secured, and only authorized users in one account can access it. Any product or negotiation process information is stored in a secure manner. Only authorized agents have access to the product information and negotiation process.
3. **No Need of Maintenance:** User can store all types of data on the cloud, and do not have to worry about maintenance of data. The product data can be stored on cloud, so that organizations do not require any server and maintenance of that server. Simultaneously maintenance cost is also reduced.
4. **No Need of Backup:** Business owner do not need to worry about the backup responsibilities, as the supplier has already taken effort to put up a great system for backup. Disk failure, server crash or system failure won't create much problem as the supplier can easily restore the latest backup from the cloud.
5. **Device and Location Independent:** User can access cloud from any device like mobile, mini-computer, and desktop etc. anywhere and anytime.

6. **Transparent Software Updates:** Softwares which are necessary with any e-commerce system, are updated, transparently and require minimum download time.

## 3   Technical Literature Survey

Li Pan [1] introduced a framework for automated service negotiation in cloud computing environments. In this framework, software agents negotiate with each other on behalf of service consumer and provider. This system also used a bilateral multi-step monotonic concession negotiation protocol for service negotiation in cloud computing environments. Service provider and consumer agents interact with each other due to the negotiation process and they make decisions according to the negotiation protocol.

Miguel A. Lopez-Carmona, Ivan Marsa-Maestre and Mark Klein [2] says that, consensus policy based mediation framework is used to perform multi-agent negotiation. This paper also proposed a mediation mechanism which is used to perform the exploration of negotiation space in the multiparty negotiation setting. The performance of mediator mechanism is under guidance of aggregation of agent performance and on the set of alternatives the mediator proposes in each negotiation round.

Mikoto Okumura, Katsuhide Fujita proposed [3], a collaborative park-design support system which is an example of collective collaboration support systems based on multi-agent systems. In this system, agents collect user information, many alternatives and reach optimal decision using automated negotiation protocol. Especially, in this paper, the attribute space and utility space of user in real world is decided. At the end of the system user gives feedback. According to the user's feedback, if most of the users agree on some alternative, then this alternative is final or optimal.

Amir Vahid Dastjerdi and Rajkumar Buyya [4] described SLA negotiation challenges in a cloud computing environment. This system also proposed time dependent negotiation which solves negotiation challenges. To increase the dependability of negotiation process, this system has included reliability assessment. Cloud providers can accommodate more requests and thus increase their profit by discriminating regarding the pattern of concession

Ivan Marsa-Maestre, Miguel A. Lopez-Carmona and Mark Klein [5] presented a framework for characterization and generation of negotiation process. Considering both the structural properties of agent utility functions, and the complexity due to relationships between utility functions of the different agents, a set of metrics to measure high-level scenario parameters is provided. Then a framework is presented to generate scenarios in a parametric and reproducible way. The basis of generator is the aggregation of hyper volumes which is used to generate utility functions. Generator is also based on the use of shared hyper volumes and nonlinear regression which is used to generate negotiation scenarios.

Bo An, Victor Lesser, David Irwin, Michael Zink [6] designed a system for dynamic resource allocation problem and implements a negotiation system. In negotiation model, multiple sellers and buyers are allowed to negotiate with each other concurrently. At the same time an agent is allowed to de-commit from an agreement at the cost of paying a penalty. This system also presents negotiation strategies for both seller and buyer.

Moustapha Tahir Ateib [7] has presented a fuzzy logic based negotiation modeling that can be used to overcome the complexity of automation negotiation processes. This system uses fuzzy logic to deal with ambiguity and uncertainties.

Yan Kong, Minjie Zhang [8] proposed a negotiation-based method which is used for task allocation under time constraints in an open, dynamic grid environment. In this environment, both consumers and provider agents can enter into or, exit the environment freely at any time. There is no central controller so that agents are negotiating with each other for task allocation based only on local views.

Hsin Rau, Chao-Wen Chen, and Wei-Jung Shiang [9] developed a negotiation model which is used for a supply chain with one supplier and one buyer. This model is useful to achieve coordination under incomplete information environment. To find an optimal solution, an objective programming approach is applied.

Liu Xiaowen, Yu Jin [10] introduced automated negotiation model for tourism industry. To improve the negotiation efficiency and success rate, this system proposed RBR and CBR. The model employs CBR method to support an automated negotiation by past successful negotiation cases used for those negotiation partners that have no contract rule existing in each other.

Mukhopadhyay et. al. recently proposed related solutions in negotiation over the Internet for efficient E-Commerce and negotiation prediction. [11] [12]

## 4 Proposed System

### 4.1 Scope of the System

Due to cloud computing, negotiation system will be secured, user can access it any time on any device like desktop, mobile etc. and organization maintenance cost is also reduced. Further we can use features like rule based reasoning and case based reasoning [10]. These two features improve the efficiency and success rate of the negotiation process.

### 4.2 Purpose

The objective of this system is to reduce maintenance cost of organizing data and provide security for data and negotiation process. So, it makes automated negotiation faster, flexible, secure and reliable.

### 4.3  Problem Definition

In order to make some agreeable decision, two or more parties come together during the negotiation process. And there are organizations to maintain data of negotiation process and product data. But this maintenance is a very tedious job. In order to overcome this problem, all organizations' product data is stored on cloud. Hence, security and maintenance cost of organizations' data is reduced.

## 5  Proposed System Architecture

In agent based negotiation system, agents are the negotiator. The negotiation process is done by agents through cloud. Cloud is used to store all product details and agents information. Further we go on case based reasoning and rule based reasoning, we will add two more databases in this system.

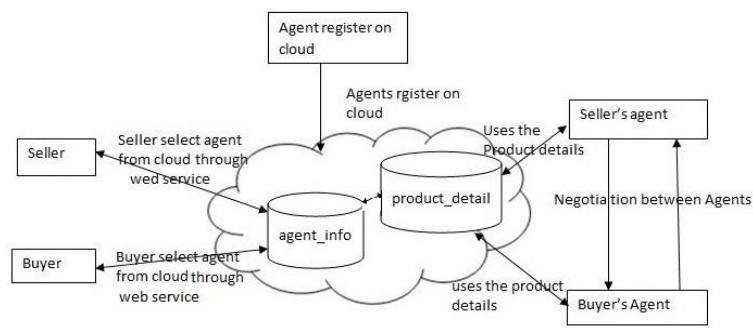

**Fig. 1.** System Architecture of Agent Based System

    If sometimes buyer has time for doing negotiation process but at the same time seller is busy in his/her work. In this situation buyer has to wait until seller is free. For this reason we can use agent based system. In this system, firstly, agents are registered on the cloud. All information related to agents (such as agent name, experience etc.) is stored on cloud database. Cloud database also has product details (for example product of company, price, features of product etc).Seller and Buyer of product selects agent using cloud database for negotiation. Agents have all requirement and details of seller and buyer respectively. Using these requirements agents negotiate. After completing negotiation, respective agents will give feedback to the seller and buyer through cloud.

For this system, we can use Amazon Simple Storage Service (Amazon S3). Amazon S3 is used as huge storage area for the internet. It is designed to make easy development of web-scale computing. Amazon S3 offers a simple web services interface which can be used to store, access and retrieve any amount of data at any time. On Amazon S3, user can write, read, and delete objects containing from 1 byte to 5 terabytes of data each. User can store unlimited number of objects.

## 5.1 System Components

For this system, we can use three modules. Using these components, system becomes easy to use and works efficiently.

1. **Store data on cloud and Agent registration:** For storing data, we can use Amazon S3 service. It makes our system fast, secure and highly reliable. In this component, agent detail and product information is stored on cloud in proper format. Only authorized agent can access product details.

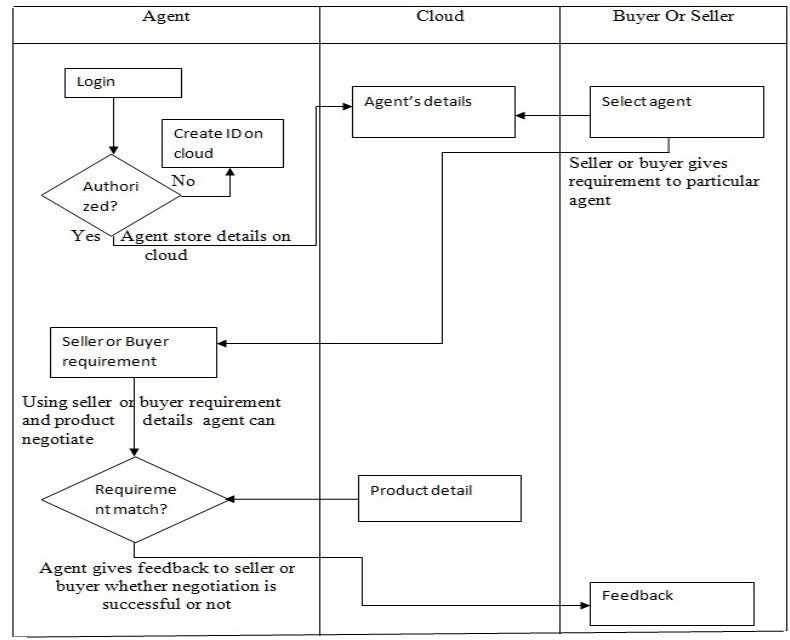

**Fig. 2.** Flowchart of the System

2. **Negotiation process:** For negotiation process, seller and buyer select their agents respectively. After that they can give their requirement to agent in encrypted format that is to generate the hash code of that requirement and encrypt that hash code using agent's public key. Agent's public key is known to sellers and buyers.

$$\text{Buyer or Seller Requirements} = E\ \{H\ (m),\ A_{pk}\}. \qquad (1)$$

Where, for generating hash function MD5 algorithm is used. $A_{pk}$ is agent's public key. We can use encryption for security purpose, in this process, we can use digital signature concept same as seller. Buyer can do same process for generating hash code of his requirement. After getting requirement, agent decrypts the hash code using his private key. And calculates the hash code for checking whether this message comes from appropriate seller or buyer and whether it is modified or not.

$$\text{Agent Receive Requirements} = D\ \{H\ (m),\ A_{pri}\}. \tag{2}$$

Where, $A_{pri}$ is agent's private key. When an agent confirms that this message comes from appropriate seller or buyer and message is not modified, then negotiation process will be start.

For negotiation process, we can use The Bilateral Negotiation Model presented by Moustapha Tahir Ateib [7].

Let x (x €{$x_1, x_2,…,x_m$}) represents the buyer agent and y(y€{$y_1,y_2,…,y_n$}) be the supplier agent. And let then i (i€{$i_1,i_2,…,i_n$}) be the issues under negotiation, such as price, volume, duration, quality and so on. Each agent assigns to each issue i, weight $W_i$ denoting the relative importance of that issue to the agent. Hence, $W_i^x$ represents the importance of issue i to agent x, therefore the overall utility function of an offer O is:

$$U(O) = \frac{\sum_{i=1}^{m} W_i u_i(x_i)}{\sum_{i=1}^{m} w_i}. \tag{3}$$

Where U(O) is the overall utility for the offer O (=[O_1,...,$O_m$] T) and $u_i(x_i)$ is the individual utility function for issue i for $u_i$ € [0,1]$ and the preference degree of an agent to an issue i is denoted as $W_i$ € [0,9]. Each agent also specifies a minimum acceptable utility level[$U_{max}, U_{min}$] to determine if an offer is acceptable. Hence, for benefit-oriented, the utility function $U_i(x_i)$ is computed as follows :

$$U_i(x_i) = \frac{x_i - x_{\{min\}}}{x_{\{max\}}^i - x_{\{min\}}^i}. \tag{4}$$

For cost oriented however, the utility function can be written as:

$$U_i(x_i) = 1 - \frac{x_i - x_{\{min\}}}{x_{\{max\}}^i - x_{\{min\}}^i}. \tag{5}$$

3. **Feedback:** When negotiation process is finished, agent gives feedback to his appropriate seller or buyer about negotiation whether it is successful or not. Then the actual E-commerce part would start which is not part of our current research.

## 6 Conclusion

Cloud computing provides various features such as security, scalability, reliability and low maintenance which are beneficial to negotiation process in E-Commerce. In this paper, we propose an agent based negotiation system, in which the agent uses cloud for storage of data and product information. Agents negotiate on some issues using the product information and seller's or buyer's requirement. After completing negotiation process, agents give feedback to user about whether the negotiation is successful or not. This negotiation system is dynamic, if number of users is increased, then number of agents also increases automatically. Our future work is to make a faster, secure and flexible E-negotiation agent using rule based reasoning and case based reasoning [10].